\documentclass[pre,twocolumn,superscriptaddress,amsmath,amssymb]{revtex4}
\usepackage{graphicx}
\usepackage{color}
\usepackage{ulem}

\begin{document}
\title{Length-dependent translocation of polymers through nanochannels}
\author{\firstname{R.} \surname{Ledesma-Aguilar}}%
\email{r.ledesmaaguilar1@physics.ox.ac.uk}
\affiliation{Rudolf Peierls Centre for Theoretical Physics, University of Oxford, 1 Keble Road, Oxford OX1 3NP, United Kingdom}
\author{\firstname{T.} \surname{Sakaue}}
\affiliation{Department of Physics, Kyushu University 33, Fukuoka 812-8581, Japan}
\affiliation{PRESTO, JST, 4-1-8 Honcho, Kawaguchi, Saitama 332-0012, Japan}
\author{\firstname{J. M.} \surname{Yeomans}}
\affiliation{Rudolf Peierls Centre for Theoretical Physics, University of Oxford, 1 Keble Road, Oxford OX1 3NP, United Kingdom}
\date{\today}

\begin{abstract}
We consider the flow-driven translocation of single polymer chains through nanochannels.  Using analytical calculations based on the de Gennes blob 
model and mesoscopic numerical simulations, we estimate the threshold flux for the translocation of chains of different number of monomers. 
The translocation of the chains is controlled by the competition between entropic and hydrodynamic effects, which set a critical penetration length 
for the chain before it can translocate through the channel. We demonstrate that the polymers show two different translocation regimes depending on how 
their length under confinement compares to the critical penetration length.  For polymer chains longer than the threshold, the translocation process is insensitive 
to the number of monomers in the chain as predicted in~Sakaue {\it et al.}, {\it Euro. Phys. Lett.}, {\bf 72} 83 (2005).   However,  for chains shorter than the critical 
length we show that the translocation process is strongly dependent on the length of the chain.  We discuss the possible relevance of our results to biological transport.

\end{abstract}
\maketitle

\section{Introduction}

The passage of polymer chains through nanochannels is an ubiquitous process in nature.  Biopolymers, such as DNA and RNA,
have to cross a multitude of barriers to perform different biological functions, for example, in translocating through cellular membrane pores or 
when ejecting from viral capsids~\cite{AlbertsCell}.  Considerable interest has arisen in the details of the translocation process due to a vast array of 
practical applications that include the  potential sequencing of DNA chains~\cite{Branton-NatBiotechnol-2008}, the sorting of biopolymers using 
smart entropic traps~\cite{Han-Science-2000} or in sieving processes~\cite{vanRijn-Nanotechnology-1998} in the pharmaceutical and food industries.   
 
The translocation of a polymer chain through a nanochannel can be described as a three-stage process: the chain must first 
find the pore, then enter it and finally move through it.   
It is the second stage, of a polymer entering a long nanochannel, that we concentrate on here.   Recent theoretical 
progress~\cite{Sakaue-EPL-2005, Sakaue-Macromol-2006, Markesteijn-SoftMatter-2009}
has shown that the entry of a polymer chain into a narrow channel driven by a fluid flow can be regarded as a tunnelling phenomenon,  
in which the entropic cost of squeezing the chain into the pore is opposed by the energy gain provided by 
the driving hydrodynamic force.  The outcome of such competition is a free energy barrier, which the polymer 
has to surpass in order that translocation takes place.

According to the de Gennes blob model for confined chains~\cite{Sakaue-EPL-2005},  
the barrier is overcome once the chain has been pushed a distance $y^*$ into the channel,  at which point, the hydrodynamic force 
wins over the entropic pressure.     Therefore, the strength and position of the the free energy barrier can be controlled by varying the 
driving volumetric flux, $J$.  
Increasing $J$ has the effect of shifting the position of the barrier closer to the channel entrance, up to a critical flux, $J_c$, at which 
the barrier height becomes comparable to the thermal energy and translocation takes place.  This gives the scaling relation
\begin{equation}
J_c \sim \frac{k_B T}{\eta},
\label{eq:Crit_Flux_Tak}
\end{equation}
where $k_B$ is Boltzmann's constant, and $T$ and $\eta$ are the temperature and the viscosity of the fluid 
medium.  

Remarkably, the threshold flux given by eq.~(\ref{eq:Crit_Flux_Tak})  is independent of the degree of polymerisation of the chain, $N$.  
Physically, this occurs because the free energy barrier is overcome when a major portion of the chain is still outside the pore.    
This behaviour thus belongs to a long-chain regime, where the polymer always reaches the position of the barrier before being 
completely brought into the channel.  Conversely, there is a short-chain regime, in which the whole chain is pushed inside the 
pore without reaching $y^*$.  In this case, the barrier still arises from the competition of the entropic pressure and the hydrodynamic 
force as before, but with the crucial difference that its magnitude and position are determined by the length of the confined chain.   
The practical consequence is that the translocation process becomes $N$-dependent, a feature of potential interest, for example,
in sorting chains according to their number of monomers.
 
In this paper we shall focus on the existence of two regimes of translocation between long and short chains, and will
examine the effect of the size of the chain on the threshold flux in each regime.  Applying the  energy barrier approach 
of Ref.~\cite{Sakaue-EPL-2005} we obtain a different scaling relation for the
threshold translocation flux for long and short chains and test our predictions 
against numerical simulations.  
  
The numerical modelling of polymer-solvent dynamics has made considerable progress during the last few years.  
In particular, coarse-grained models for the fluid that offer a physical coupling with  polymer chains have been 
developed \cite{Ahlrichs-JChemPhys-1999,Usta-JChemPhys-2005}, 
leading to a reliable representation of the dynamics of the system whilst retaining an efficient numerical performance.  
Given that the large-$N$ limit relevant to the blob model is difficult to access directly, we shall adopt
a coarse-grained representation introduced  by D\"unweg {\it et al.} \cite{Ahlrichs-JChemPhys-1999} and 
described by Usta {\it et al.}~\cite{Usta-JChemPhys-2005}, which is based on a lattice-Boltzmann 
model for the fluid coupled to a bead-spring representation of the polymer chain through an effective Stokes drag.   This approach
has been used to study the lateral migration of chains in channel flows under external forces and pressure gradients~\cite{Usta-PhysFluids-2006,
Usta-PhysRevLett-2007}.  More recently, Markesteijn {\it et al.}~\cite{Markesteijn-SoftMatter-2009} used the same model to study the 
forced translocation of long chains into narrow pores in the limit where the bead-spring model is expected to give the 
blob-model behaviour.  As expected, they confirmed the scaling of eq.~(\ref{eq:Crit_Flux_Tak}) showing that the coarse-grained model is indeed able to capture the main mechanisms at play in the
translocation process.  
   
The rest of this paper is organised as follows:  in Section~\ref{sec:Theory} we 
present scaling arguments for the position and height of the energy barrier,  and for the threshold translocation flux 
in the long-chain and short-chain regimes. 
In Section~\ref{sec:NumericalMethod} we describe the lattice-Boltzmann algorithm and the bead-spring model 
for the polymer chain, and list the set of parameters used to carry out the numerical 
simulations.  Our numerical results are presented in Section~\ref{sec:NumResults}.   After describing the simulation
setup in Section~\ref{sec:Geometry},  in Section~\ref{sec:Barrier} we study the translocation process for long chains,  
demonstrating the presence of the energy barrier.  We then focus on the cross-over to the short-chain regime in Section~\ref{sec:Cross}, 
and on the short-chain translocation process which gives rise to a length-dependent threshold flux, in Section~\ref{sec:Short}. 
Finally, in Section~\ref{sec:Conclusions} we present the discussion and conclusions of this work. 

\section{Entry of a polymer chain into a nanopore: blob model}

\label{sec:Theory}

In this section we present scaling arguments to predict the threshold flux allowing the translocation of 
linear polymer chains through a narrow channel.   As we have anticipated above, we will examine two 
regimes for the translocation corresponding to long and short chains. 

\subsection{Confined chains in equilibrium}

\begin{figure}
\centering
\includegraphics[width=0.45\textwidth]{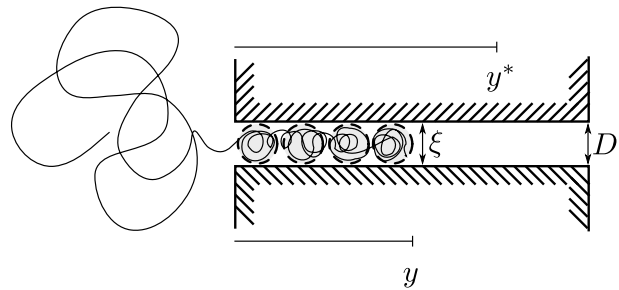}
\caption{Schematic representation of a long blob-like polymer chain under confinement.  The chain is confined up to 
a distance $y$, adopting a configuration of $M$ stacked blobs of size $\xi$.  For a 
linear chain, the size of a blob scales as $\xi \sim D$.  When subject to a driving flux, $J$, the competition between the hydrodynamic drag 
and the entropic pressure sets a barrier to translocation, located at $y^*$. Once it has reached the position of the barrier the polymer is able 
to translocate.\label{fig:DiagramR1}}
\end{figure}

We start by reviewing the scaling argument of Sakaue {\it et al.} \cite{Sakaue-EPL-2005} for 
long, blob-like linear chains.  
Consider a linear polymer chain of ideal radius $R_0\simeq aN^{1/2}$, composed of $N$ monomers of size $a$.   
In equilibrium, the Flory radius of the free chain is $R\simeq a N^{3/5}$.  

The size of the chain changes when 
it is confined in a channel of width $D<R$.   As depicted in Fig.~\ref{fig:DiagramR1}, the chain stretches 
due to the effect of confinement.  For a completely confined polymer, the equilibrium length of chain, $L$, follows 
from the minimisation of the Flory energy
\begin{equation}
\frac{F}{k_B T} \simeq \frac{L^2}{R_0^2}+\frac{N^2 a^3}{LD^2}, 
\label{eq:ConfinedFloryEnergy}
\end{equation}
and reads,
\begin{equation}
L\simeq D \left(\frac{a}{D}\right)^{5/3}N\simeq D \left(\frac{R}{D}\right)^{5/3}.
\label{eq:L}
\end{equation}
In the de~Gennes' blob picture~\cite{deGennesScaling}, the confined chain accommodates itself into $M$ blobs of uniform 
size $\xi$.    The number of blobs then obeys the relation $M=L D^2 / \xi^3.$  Within each blob the effect of confinement 
is unimportant. Hence, the size of the blob scales as $\xi \simeq a P^{3/5}$, 
where $P$ is the number of monomers in each blob.   Assuming that the blobs are stacked forming a mesh, one 
can write, for the volume fraction, 
$$\phi = \frac{Na^3}{LD^2} = \frac{Pa^3}{\xi^3},$$
from which it follows that for a linear chain, the size of the blob is comparable to the 
pore size, {\it i.e.,}
\begin{equation}
\xi \simeq D.
\label{eq:BlobSize}
\end{equation}

\subsection{Long-chain translocation}

We now consider a partly confined chain in the presence of a
driving flow.
For a weak driving flux, the conformation of the chain outside the channel is close to equilibrium.  
Hence, the translocation process is controlled by the forces acting on the confined part of the chain.  
This corresponds to the inside approach in the terminology of Sakaue {\it et al.}~\cite{Sakaue-Macromol-2006} and 
is valid as long as the length of the pore, $L_p$, is larger than its thickness, $D$~\footnote{ In the opposite limit of short channels, $L_p \lesssim D$, the 
driving force becomes localised at the position of the pore and one has to consider the resistance offered
by the whole chain in the {\it cis}  side of the duct as it moves toward the pore~\cite{Sakaue-EPJ-2006}.}.   

The confinement of the chain has an entropic penalty of the order of $k_B T$ per 
blob~\cite{deGennesScaling}.  Thus, for a chain composed of $M$ blobs which penetrates a distance $y$ into the channel, as shown in Fig.~\ref{fig:DiagramR1}, there is an energy cost
\begin{equation}
F_S \simeq k_B T \frac{y D^2}{\xi^3} = A k_B T \left(\frac{y}{D}\right),
\label{eq:FEntropic}
\end{equation}
where we have used eq.~\eqref{eq:BlobSize}, and introduced the numerical prefactor $A$ to drop the $\simeq$ symbol.
As expected, the entropic cost increases with $y$, given that a larger number of monomers are pushed into the channel.  

Countering the entropic cost is the fluid flow, 
which tends to drag the chain further into the channel. The hydrodynamic drag per 
blob scales as $ \eta u \xi$, where $u = J/D^2$ is the typical velocity in the translocation direction 
inside the channel.  For $M$ blobs, the change in free energy corresponds to the work done by the fluid to displace 
the chain up to a distance equal to $y$, 
\begin{equation}
F_H \simeq  - \eta u \int_0^y M(y') \xi(y') \mathrm d y' = - \frac{B \eta J}{2} \left(\frac{y}{D}\right)^{2},
\label{eq:FHydrodynamic}
\end{equation}
which also increases with the penetration length, both because the dragging fluid has a larger number of blobs on which 
to act, and because each blob has moved further into the channel.  As before, we introduce a proportionality constant, $B$. 

Adding the contributions given by eqs.~\eqref{eq:FEntropic} and ~\eqref{eq:FHydrodynamic}, it is possible to write the free energy change 
due to confinement under flow,  
\begin{equation}
\frac{\Delta F}{k_B T}(y) =  F_S+ F_H = A k_B T \left(\frac{y}{D}\right) - \frac{B \eta J}{2} \left(\frac{y}{D}\right)^{2}. 
\label{eq:TotFreeEnergyR1}
\end{equation}
The competition between the entropic and hydrodynamic terms gives an energy barrier, $\Delta F^*$, located at
$y = y^*$.  Differentiation of eq.~\eqref{eq:TotFreeEnergyR1} gives the position of the barrier, 
\begin{equation}
y^* = \left(\frac{A}{B}\right)\frac{k_B T}{\eta J}D,
\label{eq:BarrierR1}
\end{equation}
which can be substituted back into eq.~(\ref{eq:TotFreeEnergyR1}) to obtain its magnitude,
\begin{equation}
\frac{\Delta F^*}{k_B T} = \frac{A}{2}\left(\frac{y^*}{D}\right)=\frac{C}{2} \frac{k_B T}{\eta J},
\label{eq:FreeEnergyBarrierR1}
\end{equation}
where $C\equiv(A/B)^2B.$ 

In order that translocation proceeds, the chain must overcome this barrier.  To calculate the threshold flux 
we consider the probability of translocation of the chain,
\begin{equation}
P=\kappa_D \tau_m\exp{\left(-\frac{\Delta F^*}{k_B T}\right)},
\label{eq:Probability}
\end{equation}
where $\kappa_D \simeq k_B T/\eta D^3$ is a characteristic frequency and $\tau_m$ is the observation time of the process.   
We define the threshold flux by 
\begin{equation}
P(J_c)=P_c,
\label{eq:Pc}
\end{equation} 
where $P_c$ is an arbitrary threshold 
probability.  Inverting eq.~\eqref{eq:Pc} leads directly to the scaling 
relation~(\ref{eq:Crit_Flux_Tak}),     
$$
J_c \sim \frac{k_B T}{\eta}.
$$
  
As we had anticipated, $J_c$ does not depend on $N$. This is because the polymer reaches the position of the barrier, $y^*$,
before it is completely confined in the channel, {\it i.e.}, the translocation is triggered when $N^*<N$ monomers have 
been pushed into the channel.  To obtain the critical number of monomers, $N^*$, it suffices to replace $L$ by $y^*$ in 
eq.~(\ref{eq:L}), from which 
\begin{equation}
N^* \simeq \frac{y^*}{D}\left(\frac{D}{a}\right)^{5/3} \simeq \left(\frac{A}{B}\right) \left(\frac{D}{a}\right)^{5/3} \frac{k_B T}{\eta J},
\label{eq:CriticalN}
\end{equation}
where we have used eq.~\eqref{eq:BarrierR1}.
We note that in Ref.~\cite{Sakaue-EPL-2005} the scaling for the threshold flux was obtained by setting $\Delta F^* \simeq k_BT$ in 
eq.~(\ref{eq:FreeEnergyBarrierR1}).  This corresponds to a free energy barrier whose height is overcome by pushing 
only one blob into the pore, with a corresponding number of monomers $N_D\equiv N^*(y^*=D)\simeq (D/a)^{5/3}$.
 
\subsection{Short-chain translocation}
The short-chain regime corresponds to the limit $L<y^*$, or equivalently, to $N<N^*$, where the chain is completely confined 
{\it before} overcoming the energy barrier given by eq.~(\ref{eq:FreeEnergyBarrierR1}).  From eqs.~(\ref{eq:BarrierR1}) and~(\ref{eq:CriticalN}), 
for a given chain length, this corresponds to weak fluxes and/or wide pores. 
Instead of being located at $y^*$, the energy barrier corresponds to a penetration $y=L$, given that there is no extra 
cost for pushing the chain into the channel any further than its own length.
In terms of the number of beads, the free energy barrier obeys $\Delta F^*=\Delta F(L(N))$, and follows from 
eq.~(\ref{eq:TotFreeEnergyR1}), 
\begin{equation}
\frac{\Delta F^*}{k_B T} =  A \left(\frac{a}{D}\right)^{5/3} \left(1-\frac{1}{2}\frac{N}{N^*}\right)N,
\label{eq:FreeEnergyR3}
\end{equation}
where we have used eqs.~(\ref{eq:L}) and~(\ref{eq:CriticalN}) to write the dependence on $N$ and $N^*$ explicitly.   
According to this result, for a given constant value of the flux, the barrier can be reached more easily by decreasing the number 
of beads. Conversely, increasing $N$ has the effect of increasing the strength of the barrier, up to $N=N^*$, where one crosses 
over to the long-chain regime, and eq.~(\ref{eq:FreeEnergyR3}) reduces to eq.~(\ref{eq:FreeEnergyBarrierR1}) as expected. 

The threshold flux can be calculated from the probability of translocation, which follows after combining eqs.~\eqref{eq:Probability} and~\eqref{eq:FreeEnergyR3}.  Using the criterion given by eq.~\eqref{eq:Pc}, we obtain
\begin{align}
\frac{\eta J_c}{k_B T}  & =  2\left(\frac{A}{B}\right) \left(\frac{D}{a}\right)^{10/3}  \label{eq:CritFluxR3} \\ 
&\times  \left(\frac{\left(\frac{a}{D}\right)^{5/3}N+\frac{\log (P_c)}{A}-\frac{\log(\kappa_D\tau_m)}{A}}{N^2}\right).\nonumber
\end{align} 

The scaling \eqref{eq:CritFluxR3} should be valid in the range $N_D<N<N^*$, which corresponds to 
$D<R$ and $L<y^*$.  In this range, the threshold flux increases monotonically with the number of monomers.  
This behaviour can be traced back
to the $N$-dependent terms in eq.~\eqref{eq:FreeEnergyR3}, where the linear term, corresponding to the entropic cost,
grows faster than the hydrodynamic gain, provided that $N/N^*<1$.

\section{Numerical method}

\label{sec:NumericalMethod}
To further study the cross-over from long- to short-chain translocation discussed 
in  Section~\ref{sec:Theory}, we use the numerical method introduced in Ref.~\cite{Usta-JChemPhys-2005}.  
This is a hybrid scheme that couples a lattice-Boltzmann fluid~\cite{Ladd-JFluidMech-1994,Ladd-JStatPhys-2001} to a bead-spring model for the polymer chain.   Our work is based on the original implementation 
of the code, {\it susp3d}, which was kindly provided by the developers.  Here we present the main features of the numerical method.  For a more 
detailed description, the reader is referred to the original papers~\cite{Ladd-JFluidMech-1994,Ladd-JStatPhys-2001,Usta-JChemPhys-2005}. 

\subsection{Lattice Boltzmann Method}
\label{sec:LB}
In the lattice-Boltzmann algorithm the fluid dynamics follows from the evolution of the particle velocity distribution function ${f_i}$,  which is 
defined for a discretised velocity set $\{\vec c_i\}$.  For a given velocity vector, $f_i$ is proportional to the average 
number of particles moving in the direction of $\vec c_i$.

The dynamics of $f_i$ is given by the lattice-Boltzmann equation,
\begin{equation}
f_i(\vec r + \vec c_i\Delta t,t + \Delta t) - f_i(\vec r,t) = \Delta_i(\vec r,t) + F^{\rm ext}_i,
\label{eq:evf}
\end{equation}
which is defined over a lattice composed of nodes that are joined by the link vectors $\vec c_i \Delta t$, where $\Delta t$ is the time step
of the algorithm.  Here we use the D3Q19 model, which consists of a cubic lattice with a set of nineteen velocity vectors in three 
dimensions.  The lattice spacing, $\Delta x$, is uniform along the lattice axes. The model has three 
possible magnitudes of the velocity vectors, $\{|\vec c_i|\} = (0,1,\sqrt{2}) \Delta x/\Delta t$.  Accordingly, the $f_i$ have a 
corresponding weight $a_{c_i}$ that satisfies the condition
$\sum_i a_{c_i} = 1$.  A suitable choice of the weights for the lattice model used here is $a_0=1/3$, $a_1=1/18$ and $a_{\sqrt 2} = 1/36$.  

The dynamics expressed by eq.~(\ref{eq:evf}) is composed of two steps.  First, the distribution function undergoes a collision step, where fluid 
particles exchange momentum according to the collision operator, $\Delta_i$, and are driven by the term $F^{\rm ext}_i$, which plays the role of a body force.   
Following this collision stage, the $f_i$ are propagated to neighbouring sites in a streaming step, corresponding to the 
left-hand side of the equation.

The mapping between the lattice-Boltzmann scheme and the hydrodynamic equations follows from the definition of the hydrodynamic 
variables as moments of the $f_i$. The local density of mass, $\rho$, momentum $\rho \vec v$, and the momentum flux, $\Pi$, 
are given by 
\begin{equation}
\sum_if_i=\rho,\quad \sum_if_i \vec c_{i} =\rho \vec v,\quad {\rm and} \quad \sum_if_i \vec c_{i} \vec c_{i} = \Pi,
\end{equation}
respectively.  

In the absence of external forces, the system relaxes towards equilibrium through the collision stage in eq.~(\ref{eq:evf}).  
For the lattice-Boltzmann model used in this paper, this is done by defining the post-collision distribution function, $f_i^* = f_i + \Delta_i$, which 
can be expressed as an expansion in the fluid velocity: 
\begin{equation}
f_i^* = a_{c_i}\left(\rho + \frac{\rho \vec v \cdot \vec c_i}{c_s^2} + \frac{(\rho \vec v \vec v + \Pi^{\rm neq,*}):(\vec c_i\vec c_i -c_s^2 \mathbf{1})}{2c_s^4}\right),
\end{equation}
where $c_s=3^{-1/2}\Delta x/\Delta t$ is the speed of sound in the model. Similarly, the forcing term, $F_i^{\rm ext}$, can be expanded 
in powers of the fluid velocity (cf. Ref.~\cite{Ladd-JStatPhys-2001}). 

The post-collisional momentum flux tensor, $\Pi^{\rm neq,*}$, 
describes the relaxation towards the equilibrium momentum flux tensor, $\Pi^{\rm eq}$, according to 
\begin{equation}
\Pi^{\rm neq,*} = (1+\lambda)\bar \Pi^{\rm neq} + \frac{1}{3}(1+\lambda_\nu)(\Pi^{\rm neq}:\mathbf{1})\mathbf{1},
\end{equation}
where $\Pi^{\rm neq} = \Pi -\Pi^{\rm eq}$, $\Pi^{\rm eq}=\rho c_s^2+\rho\vec v\vec v$, and the 
bar indicates the traceless part of $\Pi^{\rm neq}$. 

The parameters $\lambda$ and $\lambda_\nu$ characterise the relaxation timescales of the lattice-Boltzmann fluid.  By performing a Chapman-Enskog 
expansion of eq.~(\ref{eq:evf}), corresponding to the limit of long length and timescales compared to the lattice spacing and the 
relaxation timescale of the fluid, the lattice-Boltzmann scheme leads to the continuity equation for the fluid density and the Navier-Stokes 
equations with second order corrections in the velocity for the fluid momentum.    In this limit the parameters $\lambda$ and $\lambda_\nu$ map to the 
shear and bulk viscosities of the fluid, 
\begin{equation}
\eta = -\rho c_s^2 \Delta t \left(\frac{1}{\lambda}+\frac{1}{2}\right),
\end{equation}  
and 
\begin{equation}
\eta_\nu = -\frac{2\rho c_s^2}{3} \Delta t \left(\frac{1}{\lambda}+\frac{1}{2}\right),
\end{equation}
respectively.  

Solid boundaries in the simulation box are implemented using the well-known bounce-back
rules \cite{Ladd-JStatPhys-2001}.  These correspond to a reflection of any
distribution function propagating to a solid node back to the fluid node it came from at the streaming stage in eq.~(\ref{eq:evf}).   
As a consequence, a stick condition for the velocity is recovered approximately halfway between the fluid node and the
solid node.

\subsection{Polymer chain}

The linear polymer is modelled by a chain composed of $N$ beads joined by freely rotating bonds.  
The bonds behave like Hookean springs, with an elastic potential 
\begin{equation}
U_{\rm el} (r) = k (r-a)^2,
\end{equation}
where $r$ is the separation between adjacent beads, $a$ is the equilibrium length of the bond 
and $k$ is the elastic constant of the spring.  

The short-range excluded-volume interactions are modelled by a truncated DLVO potential, 
\begin{equation}
U_{\rm DLVO}=U_0\frac{\exp(-\kappa_{\rm DH} r)}{r},
\end{equation}
where $\kappa_{\rm DH}$ is the Debye-H\"uckel screening length and $U_0$ is an amplitude.

The position vector of the $i$-th polymer bead, $\vec x_i$, evolves in time according to 
\begin{equation}
m\ddot {\vec{x}}_i = -\sum_{j\neq i} \vec \nabla_{ij} U + \vec F_i,
\end{equation}
where the summation term includes the excluded-volume interactions with all beads and the elastic interactions with 
neighbouring beads.  The term $\vec F_i$ contains the viscous hydrodynamic force that couples the bead 
to the lattice-Boltzmann fluid, 
\begin{equation}
\vec F_i = -\xi_0 (\vec x_i - \vec v(\vec x_i)) + \vec F_i^{\rm r}.
\end{equation}
This expression includes the Stokes drag, $\xi_0=6\pi\eta r_{\rm H},$ with $r_H$ being the 
hydrodynamic radius of the beads, and a random force $F_i^{\rm r}$ that satisfies the  
fluctuation-dissipation relation,  
\begin{equation}
\langle \vec F_i^{\rm r}(t)\vec F_i^{\rm r}(t') \rangle = 2k_B T \xi_0 \delta(t-t')\bf 1.
\end{equation}

While the polymer beads move in continuous space, the lattice-Boltzmann fluid is only defined at the lattice
nodes.  In order to calculate the coupling force, $\vec F_i$, the model uses a linear interpolation scheme to estimate
the fluid velocity, $\vec v$, at the position of the bead, $\vec x_i$. The discretisation of the lattice gives an {\it effective} hydrodynamic radius, $r_{\rm H}^{\rm eff},$ which
differs from the input hydrodynamic radius, $r_{\rm H}$.  Here we follow the same procedure as in Ref.~\cite{Usta-JChemPhys-2005}, and choose $r_{\rm H}$ in 
order to obtain the desired value of $r_{\rm H}^{\rm eff}$.    Once the hydrodynamic force has been exerted on the monomer, momentum 
conservation is enforced by exerting a force of equal magnitude back onto the fluid.

\subsection{Parameter Values}

Our objetive is to carry out numerical simulations of polymer chains in the limit where the bead-spring model
gives the blob-model behaviour presented in Section~\ref{sec:Theory}. Such regime has been validated when the  
monomer size matches the blob size, $a \simeq \xi$, for linear chains, where $\xi \simeq D$ \cite{Markesteijn-SoftMatter-2009}.  
Therefore, the bead-spring model should give the blob-limit behaviour for  $a\simeq D$.  On the other hand, the hydrodynamic radius 
of the beads, $r_{\rm H}^{\rm eff}$, is limited to small
values, corresponding to the point-particle coupling between the chain and the lattice-Boltzmann fluid~\cite{Ahlrichs-JChemPhys-1999}.  With 
these conditions in mind, we fix the bead diameter to $2r_{\rm H}^{\rm eff} = \Delta x/2$, while $a= \Delta x$ and $D=2\Delta x$.  

The remaining model parameters are chosen as follows:  we work at a fixed temperature, 
$k_B T = 0.1$.  To prevent chain crossings the spring constant is taken as $k = 300 k_B T/\Delta x^2$.
Parameter values for the excluded-volume potential are fixed to  $U_0 = k_B T\Delta x$ and $\kappa_{\rm DH} = 80/\Delta x$, 
which ensure that inter-monomer repulsions are larger than $k_B T$ at distances comparable to $r_{\rm H}^{\rm eff}$.  
To match the hydrodynamic diameter to the effective 
bead diameter, we fix $r_{\rm H} = 0.32\Delta x$ following the calibration procedure presented in Ref.~\cite{Usta-JChemPhys-2005}. 

In order to resolve the hydrodynamics correctly, one needs to ensure that the distribution function 
relaxes on a faster timescale than the diffusive timescale of the polymer chain.  This condition can be satisfied by setting 
the parameters in the collision operator in eq.~\eqref{eq:evf}  to $\lambda = -1$ and $\lambda_\nu = -1$.  
We set the fluid density to $\rho = 36$, and the timestep and lattice spacing in the lattice-Boltzmann fluid to 
$\Delta t =1$ and $\Delta x = 1$.  Using these values,  the dynamic viscosity in simulation units is $\eta = 6$.  

\section{Numerical Results}
\label{sec:NumResults}
\begin{figure}
\includegraphics[width=0.45\textwidth]{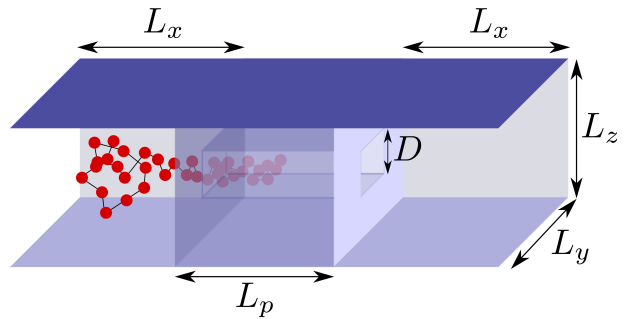}	
\caption{Schematic representation of the simulation box. \label{fig:Diagram}}
\end{figure}

\subsection{System Geometry and Initial Conditions}
\label{sec:Geometry}
The geometry of the system is depicted in Fig. \ref{fig:Diagram}. We consider two identical rectangular ducts of 
dimensions $L_x=108$, $L_y=14$ and $L_z=10$, separated by a square pore of width $D=2$ and length $L_p=24$. 
Solid walls are imposed in the $y$ and $z$ directions, while periodic boundary conditions are enforced in the $x$ direction. 

For small Reynolds numbers, one expects that the volumetric flux, $J$, caused by applying a uniform body 
force to the fluid, $f$, obeys Darcy's Law:
\begin{equation}
J = \frac{kS}{\eta}f,
\label{eq:Darcy}
\end{equation}
where $S$ and $k$ are the local cross-sectional area and permeability of the medium and $\eta$ is the  
dynamic viscosity of the fluid.  

\begin{figure}
\includegraphics[width=0.45\textwidth]{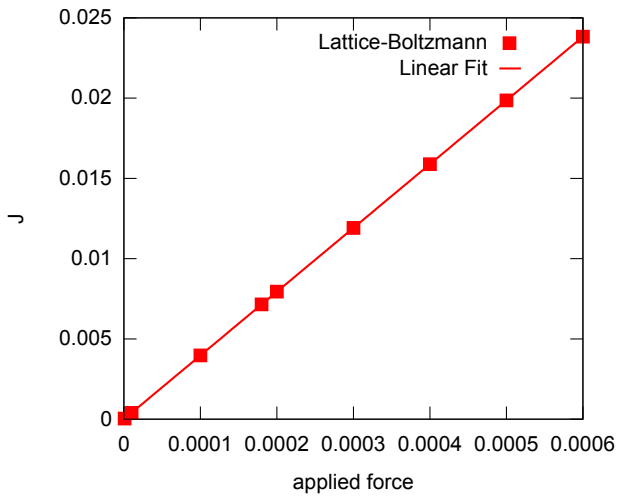}
\caption{ Volumetric flux, $J$, as a function of the body force, showing that Darcy's law holds. 
\label{fig:Darcy}}
\end{figure}

In order to verify the validity of eq.~\eqref{eq:Darcy}, we measure the local flux for different applied forcings. The results,  
illustrated in Fig.~\ref{fig:Darcy},  show a linear dependence of $J$ on $f$ as expected.  A linear fit to the data 
gives $ kS \simeq 238.$  For a given forcing, we find that the flux does not vary significantly whether it is measured
inside or outside the pore thus confirming that, for the range of forcings considered here, the lattice-Boltzmann fluid 
behaves as an incompressible liquid.  
Given that we are interested in the scaling of the critical flux, we use values of the forcing that give a ratio 
between hydrodynamic and thermal effects in the range $0<\eta J /k_BT<1.5.$  

\begin{figure}
\includegraphics[width=0.45\textwidth]{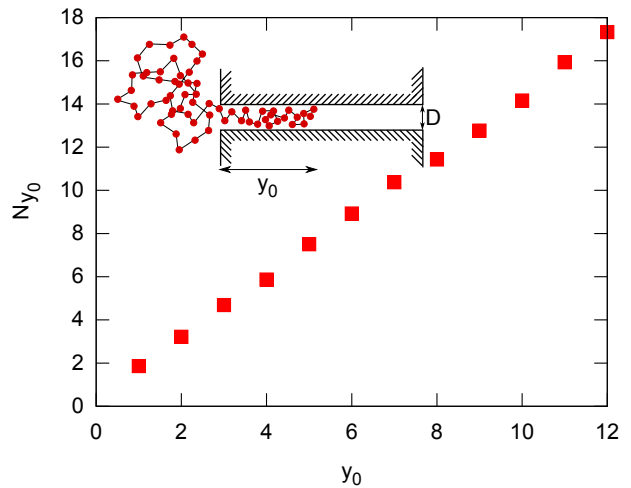}
\caption{Number of beads in the confined part of the chain as a function of the initial tethering position.  Inset: schematic representation of the tethered chain. \label{fig:Leq}}
\end{figure}

We initially tether the polymer chain at a position $y=y_0$ inside the channel as depicted in the inset of Fig.~\ref{fig:Leq}, and let the system equilibrate for 5$\times$10$^4$ time steps.   The equilibration period allows the relaxation of the chain. 
Evidence for this is presented in the same figure, which shows the number of beads inside the pore, $N_{y_0}$, as a function of $y_0$ for $N=128$.  
For a confined chain in equilibrium, we have, from eq.~\eqref{eq:L},  $N_{y_0} \simeq (y_0/D) (D/a)^{5/3}$.  Our results show that the number of 
beads increases linearly with the tethering position, as expected.  

\subsection{Energy Barrier}
\label{sec:Barrier}
In order to confirm that the polymer translocation is controlled by surpassing an energy barrier, we
first perform simulations of long polymer chains subject to a fixed driving flux while varying the initial
tethering position, $y_0$.  We consider different values of $y_0$ in the range $0 < y_0 < L_p$,  and set 
the number of monomers to $N=128$, thereby ensuring that the major part of the chain lies outside 
the pore. 

Simulations are carried out by letting the chain equilibrate as before.  Once the chain 
has equilibrated, it is released from its tethering point and the body force is applied uniformly to the 
fluid. The chain then is either carried down   the pore by the underlying fluid flow and eventually translocates to the opposite duct, or 
is ejected from the pore back into the original chamber.  Simulations are run for  $5\times10^5$ timesteps, 
which is a sufficiently long timescale to identify successful or failed translocation events.     

\begin{figure}
\includegraphics[width=0.45\textwidth]{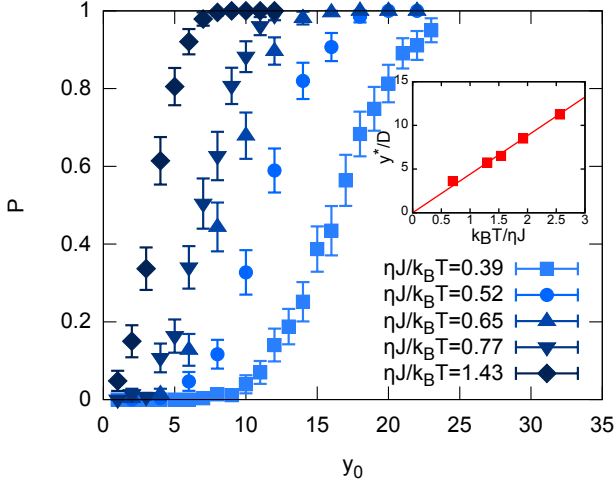} 
\caption{ Probability of translocation as a function of $y_0$.  Inset: linear growth of the position of the energy barrier with $k_BT/\eta J$. \label{fig:P.vs.y0}}
\end{figure}

Fig. \ref{fig:P.vs.y0} shows the probability of translocation of the chain as a function of $y_0$ for 
three different values of the imposed flux.  We observe a clear transition from non-translocating ($P=0$)
to translocating ($P=1$)
chains as $y_0$ is increased.  This confirms the presence of an energy barrier, which is
progressively approached as the chain is pushed further into the channel.   The probability curves
are shifted to the left as one increases the imposed flux, $J$.  This indicates a shift of the
position of the barrier, $y^*$, closer to the pore entrance caused by a higher driving hydrodynamic force.  

As a criterion we define the position of the barrier as $P(y^*)\approx 1$, and plot the measured values of $y^*$ as a function of $k_BT/\eta J$
in the inset of Fig.~~\ref{fig:P.vs.y0}.   As expected from eq.~\eqref{eq:BarrierR1}, we observe a linear growth.   
A fit of the data shown in the figure to the function $y^*(x) = (A/B) D x$ gives an estimate of the numerical
prefactor in eq.~\eqref{eq:BarrierR1},  $A/B\simeq 4.4$. 

\begin{figure}
\includegraphics[width=0.45\textwidth]{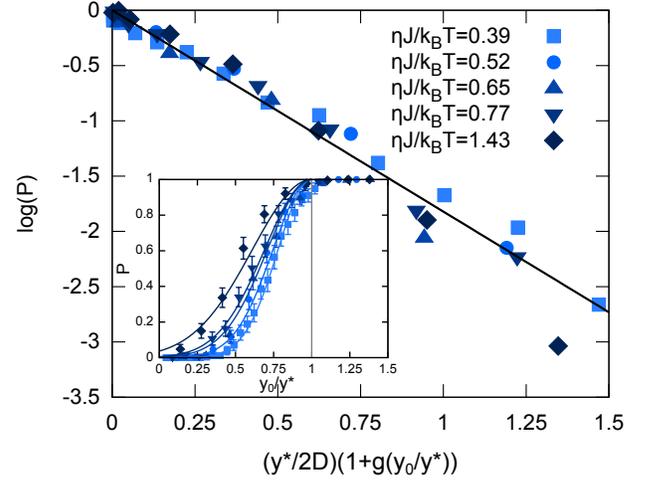}
\caption{ Logarithm of the translocation probability as a function of the free energy barrier magnitude.  
Symbols correspond to numerical simulations.  The solid line is a linear fit to the data. 
Inset: translocation probability as a function of $y_0/y^*$.   Numerical results (symbols)
show a good agreement with the theoretical prediction (solid lines).  \label{fig:P.vs.y0.collapsed}}
\end{figure}

The imposed flux sets the position and height of the energy barrier.   Given that we fix the chain 
to a prescribed position inside the channel, the height of the barrier is reduced by an amount determined by $y_0$. 
This can be easily included in the analytical model by replacing $y$ by $y-y_0$ in eq.~\eqref{eq:FEntropic} 
and performing the integration in eq.~\eqref{eq:FHydrodynamic} from 
$y'=y_0$ to $y'=y$.  The resulting free energy barrier reads 
\begin{equation}
\frac{\Delta F^*}{k_B T} = \frac{A}{2}\left( \frac{y^*}{D}\right)\left(1 + g\left(\frac{y_0}{y^*}\right)\right).
\label{eq:EnergyBarrierLongy0}
\end{equation}
where $g(x)\equiv -2x+x^2.$
From this expression and eq.~\eqref{eq:Probability} we can calculate the probability of translocation. This predicts 
 $\log P \sim \Delta F^*/k_BT$ as observed in Fig.~\ref{fig:P.vs.y0.collapsed}.  As expected, 
all data points collapse onto the same curve, which is linear in the free energy.  
A fit to the data gives an estimate for the numerical prefactor in eq.~\eqref{eq:EnergyBarrierLongy0}, 
$A \simeq 1.8$, which together with the measured value of $A/B$, gives the amplitude of the hydrodynamic contribution to the 
free energy in eq.~\eqref{eq:FHydrodynamic}, $B\simeq0.41$.   

The inset in Fig.~\ref{fig:P.vs.y0.collapsed} shows how the theoretical prediction accurately captures the main features of the translocation 
process. The probability of translocation increases to unity as  $y_0/y^*\rightarrow 1$.   The spread of the curves is dictated by the $\sim k_B T/\eta J$ prefactor
in eq.~\eqref{eq:EnergyBarrierLongy0}, increasing the likelihood of translocation at a given $y_0/y^*$ for larger fluxes, or lower temperatures.

\subsection{Cross-over from long to short chains}
\label{sec:Cross}
\begin{figure}
\includegraphics[width=0.45\textwidth]{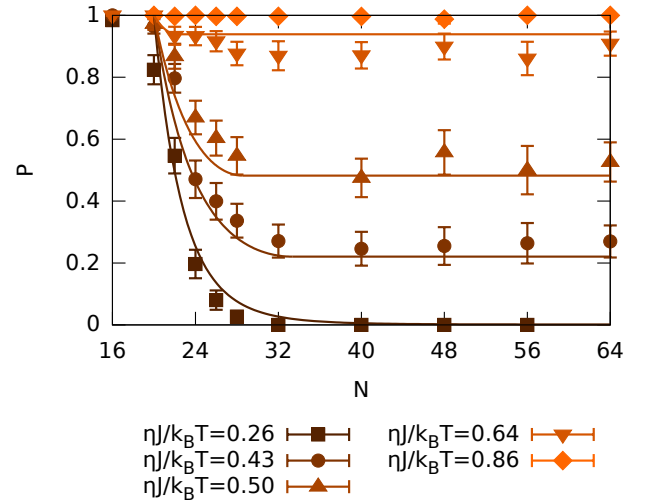}
\caption{Probability of translocation as a function of the number of beads in the chain 
at different values of the imposed flux.  Simulation results correspond to symbols, while the theoretical model is
indicated by the solid lines where we set $N_{y_0} = 20$ and $A=1.8$. \label{fig:P.vs.N.vs.highj}}
\end{figure}

So far we have discussed the translocation of long chains, which can reach the position of the barrier 
while a significant amount of monomers remain outside the pore.  The threshold flux is thus controlled
only by the number of monomers that it takes to reach the barrier, $N^*$, and not by the total number of monomers 
in the chain, $N$.

This picture breaks down for shorter chains, where our theory predicts that the scaling of the critical flux becomes 
dependent on $N$.    In order to 
explore this effect, we have carried out simulations at a fixed initial tethering position, $y_0$, 
and driving flux, while decreasing the total number of beads in the chain, $N$.      We fix the initial tethering position to $y_0=L_p/2=12,$
for which the number of beads inside the pore after equilibration is $N_{y_0}\simeq 18$.  We carry out the simulations in the range $16<N<64$, where 
we expect to observe the cross-over.  

Figure~\ref{fig:P.vs.N.vs.highj} shows the probability of translocation as a function of $N$ for five different values of the driving flux. The probability always increases with the applied
flux as expected.  For large $N$ the probability is insensitive to the chain length, over a range that persists to smaller $N$ as the flux increases.    
For smaller $N$ the probability of translocation increases strongly with decreasing $N$ until it saturates at  $N \simeq 20$  to 
$P \rightarrow 1$, where all polymer chains translocate even under the weakest flow. 

This behaviour is in agreement with our theoretical model.  For large $N$, the critical 
number of beads to overcome the barrier follows from eq.~\eqref{eq:CriticalN},  and 
obeys $N^* \sim (D/a)^{5/3} (k_BT/\eta J).$
Therefore, the cross-over to the short-chain regime ($N<N^*$) is observed at larger chain lengths
as $J$ is decreased.   From eq.~\eqref{eq:CriticalN}  we also have $N^* \sim y^*$,  from which 
it is possible to estimate the critical number of beads corresponding to each driving flux by interpolating 
the data shown in Fig.~\ref{fig:P.vs.y0}. 
Once each value of $N^*$ is known, we  can determine the energy 
barrier as a function of $N$ from eq.~\eqref{eq:FreeEnergyR3}.  
In order to include the 
effect of $y_0$, we follow the same procedure as that used to obtain eq.~\eqref{eq:EnergyBarrierLongy0}.  This gives
\begin{equation}
\frac{\Delta F^*}{k_BT} = A\left(\frac{a}{D}\right)^{5/3}\left[N\left(1-\frac{1}{2}\frac{N}{N^*}\right)+\frac{1}{2}N^*g\left(\frac{N_{y_0}}{N^*}\right)\right].
\label{eq:EnergyBarrierShorty0}
\end{equation}
Using this expression for the free energy barrier an estimate 
for the probability of translocation follows from eq.~\eqref{eq:Probability}.    
The observed saturation length $N \simeq 20$ is indeed in good agreement with our estimate $N_{y_0} \simeq 18$, at which the free energy 
barrier given by eq.~(\ref{eq:EnergyBarrierShorty0}) vanishes.  
As shown in Fig.~\ref{fig:P.vs.N.vs.highj},  we obtain a good agreement between the theory and simulations. 

\begin{figure}
\centering
\includegraphics[width=0.45\textwidth]{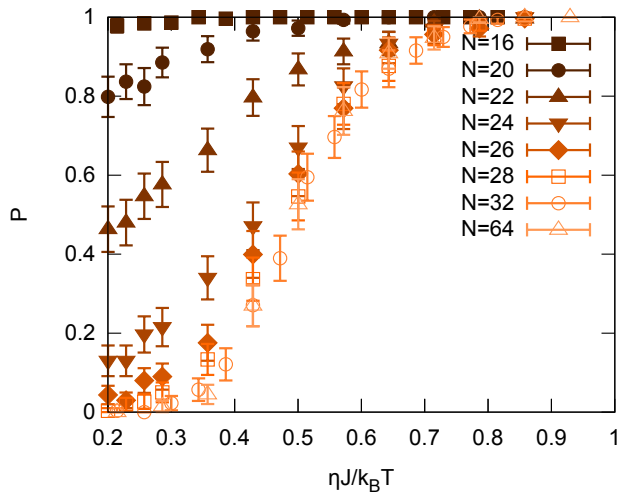}
\caption{Probability of translocation as a function of the imposed flux for different 
number of monomers in the chain. \label{fig:P.vs.j.vs.N.highj}}
\end{figure}

\subsection{Critical flux for small chains}
\label{sec:Short}
We now turn our attention to the dependence of the critical flux, $J_c$, on the number of monomers in 
the short-chain regime.  To find the threshold flux, we carry out simulations of the translocation process
at a fixed number of beads while increasing the flux.  We fix $y_0=12$ as before, and consider the range 
$0<\eta J /k_BT<1,$ for which the probability curves cross over from complete rejection of the chains to complete 
translocation.  

Figure~\ref{fig:P.vs.j.vs.N.highj} shows the resulting probability curves.  For  $N>32$ the $N$-independence of the long-chain regime is recovered, 
as curves fall on top of each other.  Our results in the long-chain regime are consistent with recent experimental studies of polymer translocation in 
nanochannels carried out by B\'eguin {\it et at.}~\cite{Beguin-SoftMatter-2011}.  In their experiments they measure the rejection coefficient of chain translocation, which is related
to the probability of translocation by ${\mathcal R}\sim 1 - P$.  Their experiments give the same smooth transition from 
rejection to translocation of the chains in the range $0<\eta J /k_BT<1$, very close to our simulation results.  For $N<32$ the probability curves shift upwards and 
become increasingly plateau-like as the number of monomers is decreased, indicating the cross-over to the short-chain regime.     

\begin{figure}
\includegraphics[width=0.45\textwidth]{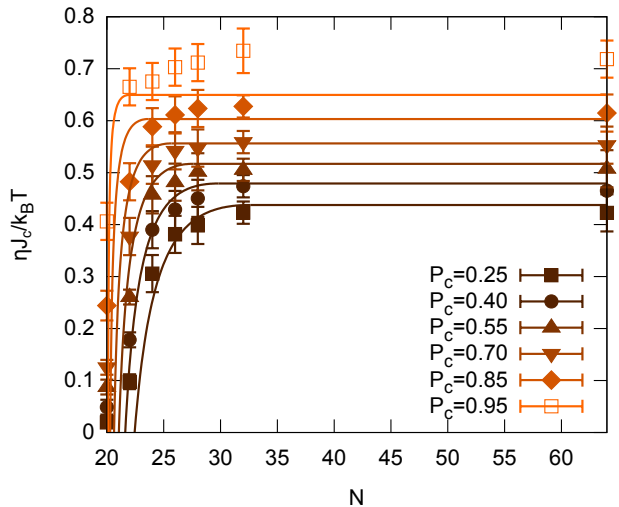}
\caption{Threshold flux as a function of the number of monomers in the chain.  $P_c$ is the choice of threshold probability. Simulation results: symbols.  
Theoretical prediction: solid lines. 
\label{fig:jc.vs.N.highj}}
\end{figure}

In order to quantify the cross-over from the short- to the long-chain regime, we use the criterion given by eq.~\eqref{eq:Pc} to interpolate $J_c$ from each of the curves shown in Fig.~\ref{fig:P.vs.j.vs.N.highj}.  
The threshold flux is shown in Fig.~\ref{fig:jc.vs.N.highj}
as a function of $N$ for six different values of the threshold probability.  As expected, the overall behaviour 
does not depend on the particular choice of $P_c$.   The threshold flux  saturates for large $N$, corresponding to the long-chain regime, 
and shows a marked decrease as $N\rightarrow N_{y_0}$, where the chain becomes completely confined inside the channel. 

Using eqs.~\eqref{eq:Probability},~\eqref{eq:Pc}  and~\eqref{eq:EnergyBarrierShorty0}, we can estimate the threshold flux 
as a function of the number of monomers in the chain,  
\begin{align}
\frac{\eta J_c}{k_B T} & \simeq  2\frac{A}{B} \left(\frac{D}{a}\right)^{10/3}\label{eq:ThersholdFluxShorty0}\\ 
&\times \left(\frac{\left(\frac{a}{D}\right)^{5/3}(N-N_{y_0})+\frac{\log (P_c)}{A}-\frac{\log(\kappa_D\tau_m)}{A}}{N^2-N_{y_0}^2}\right).  \nonumber
\end{align}

Both prefactors, $A$ and $B$, depend on the intrinsic properties of the chain and not on the particular translocation regime. We therefore set them to the values 
that we have measured previously.  
 
Figure~\ref{fig:jc.vs.N.highj} shows the resulting comparison between theory and simulation.     
Remarkably, we find a good quantitative agreement by using no new fit parameters 
in our theoretical prediction.  
\section{Summary and Discussion}

\label{sec:Conclusions}
The results presented in this paper  show that the translocation of polymer chains 
through narrow pores exhibits two different regimes depending on the length of the chain.  
As previously proposed in Ref.~\cite{Sakaue-EPL-2005}, 
our numerical results show that the translocation process is controlled by overcoming a free energy barrier characterised by a critical penetration 
of the polymer inside the pore, $y^*$.  For long chains, this length scale always outruns the polymer length under confinement, $L$.  Thus, the 
translocation process is controlled by $y^*$, regardless 
of the number of monomers in the chain, $N$.    Conversely, for small chains, where the confined length of the polymer is smaller than $y^*$, 
we have shown that the translocation process is controlled by  $L$, and the process  becomes $N$-dependent.  A qualitatively similar 
conclusion should apply to the passage of the chains through two dimensional narrow slits~\cite{Sakaue-EPJ-2006}. 

For even smaller chains, where 
$L$ is comparable to the size of the pore, $D$, and for very small fluxes ($\eta J /k_B T\simeq 10^{-1}$), we 
have observed a reduction of the probability of translocation of the chains.  This is caused by the effect of diffusion of the 
chain inside the pore, which can favour the ejection of small chains at very low forcings. However, this regime falls out of the 
free energy barrier picture presented in this paper, and we therefore leave it for future exploration.   

An application of the dependence of the translocation probability of short chains on $N$ is the potential sorting of 
the chains according to their number of monomers. In the short-chain regime the polymer must be completely 
pushed in before translocating through the pore. Given that only a small number of blobs can be pushed into the pore by fluctuations, 
one expects that only very short chains translocate by virtue of thermal effects in this regime.   For longer chains, but still in the short-chain
regime, the translocation must be assisted,  for example, by a molecular motor that is able to push the chain deeper into the pore.   

In an aqueous solution at room temperature, the critical flux in the long chain regime follows the scaling of eq.~\eqref{eq:Crit_Flux_Tak} and is 
of order $J_c\sim 10^{-18}~{\rm m^3~s^{-1}}$.  For a micrometre sized channel the corresponding velocity, $v_c\simeq J_c /D^2$, is of order 
$10^{-6}~{\rm m~s^{-1}}$.   Such a small value suggests that chains can easily translocate through microfluidic chambers at normal operation
velocities, which are typically of centimetres to metres per second.  However, recent experimental measurements of cytoplasmic streaming in 
micrometre cell channels~\cite{vandeMeent-JFluidMech-2009} show that the streaming velocities are in the range of $10~{\rm \mu m ~s^{-1}}.$  
For weaker fluxes than $J_c$, one enters the short-chain translocation regime.  Therefore, it is feasible
that the translocation regime proposed in this paper might be at play in biological systems as a regulator of selective translocation. 

Here we have considered polymers in a good solvent using the 
DLVO potential to implement short-ranged excluded-volume interactions with the wall. 
Understanding the details of electrostatic effects, with regards both to the interaction with the wall and the interactions
between segments, and how they are affected by the salt concentration, is an important issue for future work, relevant, in particular, to the dynamics 
of DNA in practical applications.  

Finally, we comment on the feasibility of an experimental confirmation of our prediction.   B\'eguin {\it et at.}~\cite{Beguin-SoftMatter-2011} have recently performed an experimental study of the flow-driven translocation of hydrosoluble polymers into nanopores.  
In their experiments, they consider chains whose radius of 
gyration is $R\approx84$~nm, and which are forced across pores $23-35$~nm in radius and $6.5~\mu$m in length.  Their results for the rejection 
coefficient,  $\mathcal R= 1 - \exp{\left(\Delta F^*/k_BT\right)},$ show a good agreement with the long-chain limit of the de~Gennes model, where they use a 
similar expression to eq.~\eqref{eq:FreeEnergyBarrierR1}.  According to our prediction, the short chain regime would be observable for chains
whose length under confinement is smaller than the position of the barrier.  This traduces into a cross-over radius of gyration 
$R^*\simeq D (k_BT/\eta J)^{3/5}$. Taking $D=70~$nm and $\eta J/k_BT \approx 0.8$, which correspond to experimental conditions reported
in ref.~\cite{Beguin-SoftMatter-2011}, we estimate $R^*\approx80~$nm, which is close to the radius of gyration of the polymers used in the
experiments.  This supports the feasibility of future experimental work to verify the theoretical predictions presented in this paper.
 
\section{Acknowledgements}
We kindly thank Berk Usta and Tony Ladd for providing us with a copy of the numerical code.  RL-A wishes to thank A. Chauduri for useful discussions. 
RL-A and JMY acknowledge support from EPSRC Grant EP/D050952/1. 
 
\providecommand*{\mcitethebibliography}{\thebibliography}
\csname @ifundefined\endcsname{endmcitethebibliography}
{\let\endmcitethebibliography\endthebibliography}{}

\end{document}